\documentclass[aps,prl,superscriptaddress,preprintnumbers,twocolumn,groupedaddress,showpacs]{revtex4}
\usepackage{txfonts}
\usepackage{graphicx}
\usepackage{slashed}
\usepackage{epsfig}

\begin{document}

\newcommand{\pphh}{$pp\rightarrow HH~$}

\title{Constraints on Large-Extra-Dimensions model through 125GeV Higgs pair production at the LHC}
\author{
Sun Hao$^{1}$, Zhou Ya-Jin$^{2}$ and Chen He$^{3}$\\
{\small $^{1}$ School of Physics and Technology, University of
Jinan, Jinan 250022, Shandong Province, P.R.China} \\
{\small $^{2}$ School of Physics, Shandong University, Jinan 250100,
Shandong Province, P.R.China} \\
{\small $^{3}$ Guangdong University of Foreign Studies, Guangzhou
510420, Guangdong Province, P.R.China } \\ }



\begin{abstract}
Based on the analysis of 5 $\text{fb}^{-1}$ of data at the LHC, the
ATLAS and CMS collaborations have presented evidence for a Higgs
boson with a mass in the 125 GeV range. We consider the 125 GeV
neutral Higgs pair production process in the context of
large-extra-dimensions(LED) model including the Kaluza-Klein(KK)
excited gravitons at the LHC. We consider the standard model(SM)
Higgs pair production in gluon-gluon fusion channel and pure LED
effects through graviton exchange as well as their interferences. It
is shown that such interferences should be included, the LED model
raises the transverse momentum( $P_t$) and invariant mass($M_{HH}$)
distributions at high scales of $P_t$ and $M_{HH}$ of the Higgs pair
production. By using the Higgs pair production we could set the
discovery limit on the cutoff scale $M_S$ up to 6 TeV for $\delta=2$
and 4.5 TeV for $\delta=6$.
\end{abstract}

 \maketitle

\par
Gravitation is by far the weakest force of nature with the usual
explanation that quantum gravitational effect only become important
at the Planck mass scale $M_{pl}=G_N^{-1/2}=1.22\times10^{19}$ GeV.
The fact that this scale is so much higher than the SM electroweak
scale of {$\cal O $}(100 GeV ) leads to the hierarchy problem.
Arkani-Hamed, Dimopoulos, and Dvali (ADD)\cite{ADD:1}\cite{ADD:2}
proposed that extra spatial dimensions could potentially solve such
problem in the SM\cite{ADD:solve}. They proposed a scenario in which
the SM is constrained to the common 3+1 space-time dimensions
(brane), while gravity is free to propagate throughout a larger
multidimensional space (bulk). The gravitational flux on the brane
is therefore diluted by virtue of Gauss's Law in the bulk, which
relates the fundamental Planck scale $M_S$ to the apparent reduced
scale according to the formula $M^2_{pl}=8\pi M^{\delta+2}_{S}
R^\delta$ , where $R$ and $\delta$ are the size and number of the
extra dimensions(ED), respectively. Postulating a fundamental Planck
scale to be on the order of the electroweak symmetry breaking scale
(1 TeV) results in ED with $R < 1$ mm for $\delta \geqslant 2$.

\par
At hadron colliders, the pair production of Higgs bosons plays a
distinctive role in understanding the Higgs mechanism\cite{Higgs}.
As the triple self-coupling of Higgs particles is involved in such
production thus provide the experimental reconstruction of the Higgs
potential. Precise measurement of this coupling could therefore give
more insight on the mechanism of electroweak symmetry breaking.
Further more, compared to that of a single Higgs boson production,
the signal-to-background ratio could significantly improved. The
invariant mass scale of the single Higgs production is fixed by the
Higgs mass, of order only $\sim$ 100 GeV. Thus their detection
through heavy quark decay modes suffer from large QCD backgrounds.
Besides, one viable decay mode $h\rightarrow\gamma\gamma$ has a very
small branching ratio of order $10^{-3}$\cite{Hrr}. While for the
pair production of the Higgs particles, the four b-jets in the final
states are energetic and reduce main background coming from $Hb{\bar
b}$ with soft b-jets\cite{Hbb}, thus provide better
signal-to-background ratio.

\par
Another important distinctive feature of the Higgs pair production
at the LHC is that the effects of physics beyond the SM can
remarkably enhance the cross section with respect to that of the SM.
Phenomenological studies of Higgs pair production have thus been
performed in the context of the 4th generation
model\cite{sunhaopphh}, the littlest higgs model\cite{wangleipphhLH}
and the Universal Extra Dimensions model\cite{pphhUED}. For the
large extra dimensional models, the tree level diagrams mediated by
the Kaluza-Klein gravitons lead to a large total cross section. Such
new theoretical approaches have drawn extensive attention in
ref\cite{pphhED} for a comparantion between supersymmetry and LED
models. In the context of the large-extra-dimensions(LED) model,
lower limits are set on the effective Planck scale in the range of
2.3$\sim$3.8 TeV at the 95$\%$ confidence level\cite{LEDbounds}.
These limits are the most restrictive bounds on virtual graviton
exchange to date. Based on the analysis of 5 $fb^{-1}$ of data at
the LHC, the ATLAS\cite{SMHiggs125GeV_ATLAS} and
CMS\cite{SMHiggs125GeV_CMS} collaborations have presented evidence
for a Higgs boson with a mass in the 125 GeV range. We thus
concentrate on the 125 GeV Higgs pair production related to the
latest measurement with the effects of the LED models and find the
characteristic distribution of it and give the observable and
unobservable limits of the effective Planck scale theoretically.

\begin{figure}[hbtp]
\vspace{-3cm} \hspace*{-4.8cm}
\includegraphics[scale=0.7]{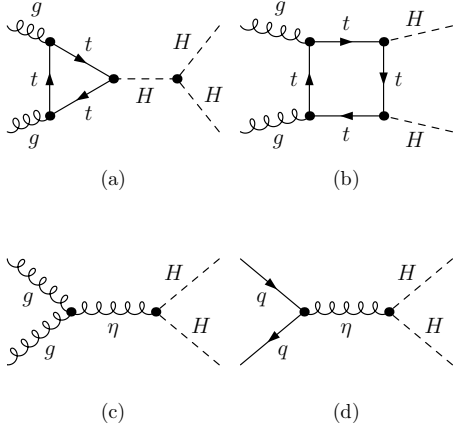}
\vspace{-12cm} \caption{\label{fig1} Part of Feynman diagrams for
\pphh in the SM and LED, where $q$ represents u-,d-,c-,s-, and
b-quark while $\eta$ represents spin-2 KK graviton.}
\end{figure}

\par
In FIG.\ref{fig1} we display the Feynman diagrams for this rare
production \pphh at the LHC in both the SM and LED model. There is
gluon-gluon fusion channel contribute to SM predictions through
top-quark loops, see FIG.\ref{fig1}(a,b) for more details. There is
also b-quark contribution, but it is small. Other diagrams include
the change of the loop arrow in FIG.\ref{fig1}(a,b) and cross change
of the legs in FIG.1(b) which are similar thus not shown, totally
eight Feynman diagrams contribute. Contributions from the
quark-antiquark collision can be safely omitted in the light fermion
mass limits except b-b fusion through t-channels. However, it is
only less of 0.5 percent of gluon-gluon fusion contribution, thus
not considered here\cite{sunhaopphh}. Now let's study the effects of
KK excitation of gravitons on the production cross section of \pphh.
There exist tree level Feynman diagrams mediated by
KK-gravitons($\eta$) only in s-channel: the gluon-gluon
fusion(FIG.\ref{fig1} (c)) and quark-antiquark
collision(FIG.\ref{fig1} (d)). Furthermore, gravitons with
polarizations that lie entirely within the physical dimensions are
effective spin 2 objects. So we will primarily be concerned with the
effects of the exchange of virtual spin 2 gravitons. To perform
perturbative calculations in this theory, one can formulate Feynman
rules for the couplings of graviton states to ordinary particles.
Related couplings in FIG.1(c,d) can be found in ref.\cite{ADD:Gian,
ADD:HanTao}. Such couplings would proportional to
$\kappa=\sqrt{16\pi G_N}$ which is the effective expansion
parameter. In particular, we adopt the conventions of
ref.\cite{ADD:HanTao}. We define the SM cross section as
$\sigma_{SM}$ and use $\sigma_{LED}$ to perform pure LED cross
section plus the interference between SM and LED effects.
$\sigma_{tot}$ is defined as $\sigma_{SM}+\sigma_{LED}$.

\par
In the case of the exchange of virtual graviton states, one must add
coherently the effect of each graviton. For instance, in our case of
only s-channel exchange, the propagator is proportional to
$i/(s-M^2_{\eta})$ where $M_{\eta}$ is the mass of the graviton
state $\eta$. Thus, when the effects of all the gravitons are taken
together, the amplitude is proportional to $\sum_{\Lambda}
\frac{i}{s-M^2_{\eta}}=D(s)$. If $\delta \geq 2$ this sum is
formally divergent as $M_{\eta}$ becomes large. We assume that the
distribution has a cutoff at $\Lambda \sim M_S$, where the
underlying theory becomes manifest. After summing over all KK
states, the effective graviton propagator $D(s)$ times square of the
coupling $\kappa$ can be expressed as
\begin{eqnarray}
\kappa^2 D_s =
\frac{8\pi}{M_S^4}(\frac{\sqrt{s}}{M_S})^{\delta-2}[\pi + 2i
I(\Lambda/\sqrt{s})].
\end{eqnarray}
The function I($\Lambda/\sqrt{s}$) depends on the ultraviolet cutoff
$\Lambda$ on the KK modes and its expression can be found in
ref.\cite{ADD:HanTao}. The default choice for $\Lambda$ would be the
fundamental scale $M_s$.

\par
We use FeynArts, FormCalc and LoopTools
package\cite{FeynArts,FormCalc,LoopTools} to perform the numerical
calculation. We use BASES\cite{BASES} to perform the phase space
integration and cern libary to display the distributions. In the
numerical calculations, we take the input as $M_Z=91.1876~{\rm
GeV}$, $M_W=80.399~{\rm GeV}$,
$\alpha(0)^{-1}=1/137.035999779$\cite{ParticleDataGroup}. The
factorization scale is chosen to be $\mu=M_{H}=125$ GeV at 14 TeV
LHC with the luminosity $\cal {L}$ as a running parameter. Typically
the latest new parton distributions for collider physics
CT10\cite{CT10} has been used in our calculation.

\begin{figure}[hbtp]
\centering
\includegraphics[scale=0.4]{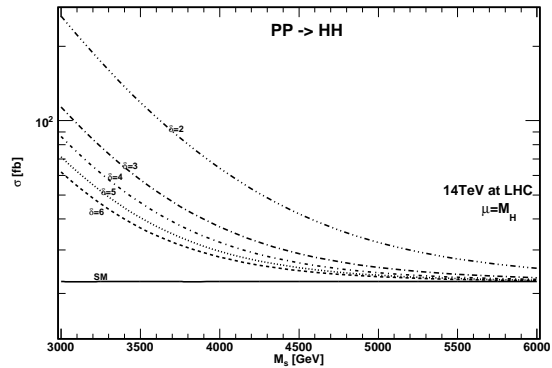}
\caption{\label{fig2} The cross sections for the process \pphh in
the SM and LED model as functions of $M_S$ with $\mu=M_H$ and
$\delta=2,3,4,5,6$ at the $\sqrt{s}=14 TeV LHC$. }
\end{figure}

\begin{figure}[hbtp]
\centering
\includegraphics[scale=0.4]{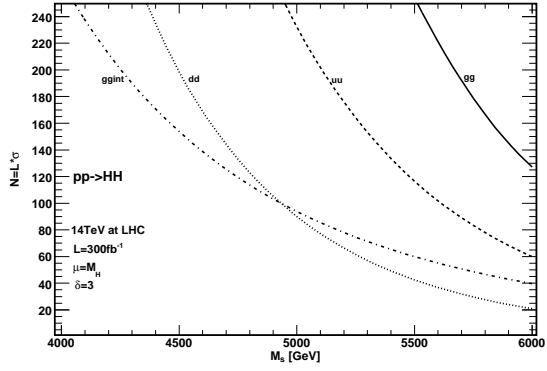}
\vspace*{-0.5cm} \caption{\label{fig3} The expected number of events
produced of separate contributions of LED effects for \pphh with
$M_s=3 TeV$, $\mu=M_H=125 GeV$ and $\delta=3$ at 14 TeV LHC . The
luminosity is taken as ${\cal L}=300fb^{-1}$. Here, gg, uu, dd and
ggint present the gluon-gluon fusion, up-quark collision, down-quark
collision and interference contributions.}
\end{figure}

\par
In FIG.\ref{fig2}, we depict the cross sections for the process
\pphh at 14 TeV LHC in the LED model as the functions of the
fundamental scale $M_S$ from 3 TeV to 6 TeV, with $\mu=M_H$ and the
extra dimension number $\delta$ being 2,3,4,5 and 6, respectively.
The solid line presents the SM prediction $\sigma_{SM}$ and the
other lines correspond to $\sigma_{tot}=\sigma_{SM}+\sigma_{LED}$.
From the figures one finds that the largest deviation from the SM
due to LED occurs at small values of $M_S$ and $\delta$. In order to
compare the relative size of different sub-contributions, we display
in FIG.\ref{fig3}, the expected number of events(${\cal N}={\cal
L}\cdot\sigma$) for each sub-contribution to Higgs pair production
for $\delta=3$, $\mu=M_H$ and the luminosity ${\cal L} =
300fb^{-1}$. The terms gg, uu, dd, ggint refer to pure gluon-gluon
fusion, $u\bar{u}$, $d\bar{d}$ collisions and to the gluon-gluon
interference term respectively. As can be seen, as the scale $M_S$
becomes larger, all sub-contributions turn out to be of the same
order. Moreover, the gluon-gluon interference term can be even lager
than the down-quark contribution when $M_S$ is larger than 4.9 TeV.
Thus all these contributions should be included for a precise
prediction.

\begin{figure}[hbtp]
\centering
\includegraphics[scale=0.4]{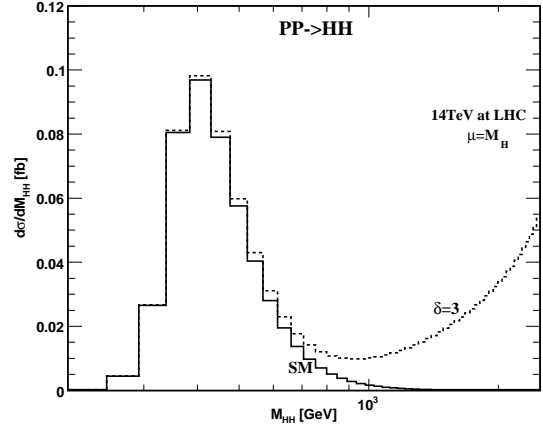}
\includegraphics[scale=0.4]{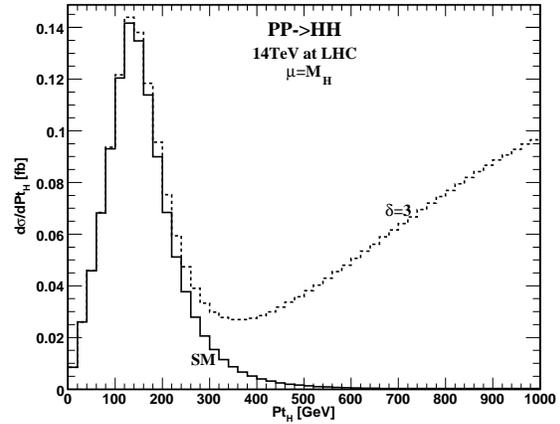}
\includegraphics[scale=0.4]{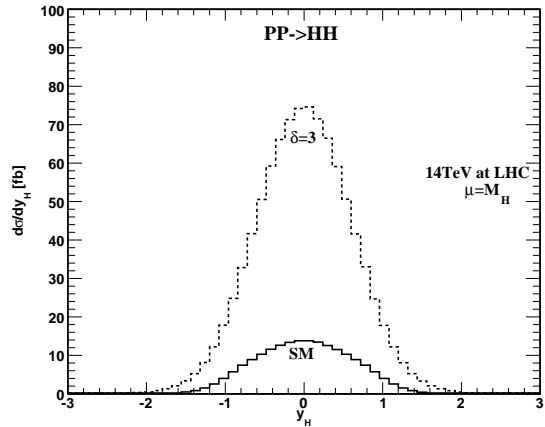}
\caption{\label{fig4} The invariance mass $M_{HH}$,  transverse
momentum $P_t$ and rapidity $\eta$ distribution with $M_s=3 TeV$,
$\mu=M_H=125 GeV$ and $\delta=3$ at 14 TeV LHC for $pp\rightarrow
HH$. The first, second and third figure correspond to (a),(b) and
(c), separately. }
\end{figure}

\par
The distributions of the Higgs pair invariance mass $M_{HH}$ and the
Higgs boson transverse momentum $P_t$ as well as the rapidity $\eta$
at 14 TeV LHC, are shown in Fig.\ref{fig4} (a), (b) and (c),
separately. There the results are for $M_s=3$ TeV, $\mu=M_H=125$ GeV
at the fixed value 3 for the number of extra dimensions and obtained
by taking the input parameters mentioned above. In FIG.\ref{fig4}
(a), (b), the SM and LED invariance mass peaks around the threshold
$\sqrt{s}= 250 $ GeV while $P_t$ distribution peaks around 150 GeV.
The LED effects gently raise the $M_{HH}$ and $P_t$ distributions at
values of high $M_{HH}$ and $P_t$ regions. Rapidity distribution is
defined as $\frac{d\sigma}{d\eta}$ with $\eta=\frac{1}{2}
\textbf{ln}(P_1\cdot q)/(P_2\cdot q)$, where $P_1$ and $P_2$ are
incoming proton momenta and q is the sum of the Higgs boson
4-momenta. As we can see, the rapidity distributions in the LED
model show significantly narrow peaks around $\eta = 0$, which
implies the large contributions at high $P_t$ region.

\par
It is clear that if the deviation of the cross section from the SM
prediction is large enough, the LED effects can be found. We assume
that the LED effects can and cannot be observed, only
if\cite{sunhaorrttH}
\begin{eqnarray}
|\sigma_{LED}|\geq  \frac{5\sqrt{ {\cal L}\sigma_{tot}}}{{\cal L}}
\equiv 5\sigma
\end{eqnarray}
and
\begin{eqnarray}
|\sigma_{LED}|\leq  \frac{3\sqrt{ {\cal L}\sigma_{tot}}}{{\cal L}}
\equiv 3\sigma
\end{eqnarray}
,respectively. In TABLE.1, we present the discovery and exclusion
fundamental scale $M_S$ values at the 14 TeV LHC with the luminosity
100$fb^{-1}$, 200$fb^{-1}$ and 300$fb^{-1}$. It shows that by using
the Higgs pair production we could set the discovery limit on the
cutoff scale $M_S$ up to about 6 TeV for $\delta=2$ and 4.3 TeV for
$\delta=6$ with the luminosity ${\cal L}=300fb^{-1}$ at the LHC.
Other similar analysis of pair productions that probe the LED model
through virtual effects of KK modes have been studied in processes
like di-lepton, di-gauge boson($\gamma\gamma$, $ZZ$, $W^+W^-$),
$t\bar{t}$ pair,
dijet\cite{ADDvirtualA,ADDvirtualB,ADDvirtualC,ADDvirtualD} etc. At
LHC, it is expected that $t\bar{t}$ production can be used to
explore a range of $M_S$ values up to 4 TeV. Through di-photon
production the LHC can extend this search to 5.3-6.7 TeV, depending
on the number of extra dimensions. While using di-lepton production,
a lower bound of $M_S$ in the 6.5 to 12.8 TeV range can be obtained.
The phenomenology of the Higgs pair production is at the time much
richer\cite{HiggsPairImp}, though Higgs pair production cannot give
compete limits as, for example, di-lepton production gives, it's
still very important.

\begin{table}
\begin{center}
\begin{tabular}{l c c c c c r r r r r r }
\hline\hline
  ${\cal L}$    &&\multicolumn{2}{c}{100$fb^{-1}$}&& \multicolumn{2}{c}{200$fb^{-1}$}&& \multicolumn{2}{c}{300$fb^{-1}$} \\ [0.5ex]
      && $5\sigma$ & $3\sigma$&& $5\sigma$ & $3\sigma$ && $5\sigma$ & $3\sigma$ &&  \\
\hline
$\delta=2$ && 5396 & 6121  && 5720 & 6488 && 5917 & 6699  \\
$\delta=3$ && 4568 & 5128  && 4824 & 5405 && 4978 & 5551  \\
$\delta=4$ && 4294 & 4810  && 4528 & 5046 && 4666 & 5193  \\
$\delta=5$ && 4113 & 4612  && 4333 & 4822 && 4469 & 4962  \\
$\delta=6$ && 3983 & 4458  && 4199 & 4674 && 4312 & 4789  \\
\hline\hline
\end{tabular}
\end{center}
\vspace*{-0.8cm}
\begin{center}
\begin{minipage}{7cm}
\caption{\label{tab3} The discovery ($|\sigma_{LED}| \geq 5\sigma$)
and exclusion ($|\sigma_{LED}| \leq 3\sigma$) LED model fundamental
scale ($M_S$) values for the $pp\rightarrow HH$ processes at the
$\sqrt{s}$ = 14 TeV LHC.}
\end{minipage}
\end{center}
\end{table}

\par
In a short summary, we calculate the \pphh process in the SM and LED
model at the 14 TeV LHC with the 125 GeV mass Higgs pair production.
We keep all the contributions include gluon-gluon fusion,
quark-antiquark collisions as well as gluon-gluon fusion
interference terms between SM and LED. We investigate the integrated
cross sections, the distributions of some kinematic variables
$M_{HH}$, $P_t$ and $\eta$. The 5$\sigma$ discovery and 3$\sigma$
exclusion ranges for the LED parameters $M_S$ are also obtained. The
effects of the virtual KK graviton turns out to enhance the
differential distributions of kinematical observables generally.
With the observable of 125 GeV Higgs boson production, more
information related to LED effects can be obtained experimentally
through such important production.

\vspace*{0.1cm} \vskip 1mm
\par
\noindent{\large\bf Acknowledgments:} We would like to thank Fawzi
Boudjema to read the paper. Project supported by the National
Natural Science Foundation of China (Grant No.11147151,
No.11105083 and No.10947139).

\end{document}